\documentclass{sf2a-conf2011}
\usepackage{graphicx}
\usepackage{hyperref}
\usepackage[]{natbib}  
\usepackage[cyr]{aeguill}
\usepackage{epstopdf}

\def\BibTeX{{\rm B\kern-.05em{\sc i\kern-.025em b}\kern-.08em
    T\kern-.1667em\lower.7ex\hbox{E}\kern-.125emX}}
\bibpunct{(}{)}{;}{a}{}{,}  

\begin{document}

\TitreGlobal{SF2A 2011}


\title{Lepto-hadronic modelling of blazar emission}

\runningtitle{Lepto-hadronic modelling of blazar emission}

\author{M. Cerruti}\address{Laboratoire Univers et Théories (LUTH), Observatoire de Paris, CNRS, Université Paris Diderot, Meudon, France}
\author{A. Zech$^1$}
\author{C. Boisson$^1$}
\author{S. Inoue}\address{Institute for Cosmic Ray Research, University of Tokyo, Kashiwa, Japan}



\setcounter{page}{237}

\index{Cerruti, M.}
\index{Zech, A.}
\index{Boisson, C.}
\index{Ionue, S.}


\maketitle


\begin{abstract}
 The characteristic double-bumped spectral energy distribution (SED) of blazars is explained by either leptonic or hadronic models. In the former, Inverse Compton emission dominates the emission of the high energy bump, while proton synchrotron emission and proton-gamma interactions dominate it in the latter.\\
  We present a new stationary lepto-hadronic code that evaluates both the leptonic and the hadronic interactions. Apart from the modelling of the SED produced in a leptonic or hadronic model, the code permits the study of interesting mixed lepto-hadronic scenarios, where both processes contribute significantly to the high energy bump.\\
  A first application to data from the high frequency peaked BL Lac object PKS~2155-304 is discussed.\\ 
\end{abstract}

\begin{keywords}
Radiative transfer, BL Lacertae objects, Gamma rays
\end{keywords}


\section{Introduction}
 The spectral energy distribution (SED) of blazars is characterised by two non-thermal emission bumps, the first one peaking in the optical or in X-rays, the second one peaking in $\gamma$-rays. 
The origin of this emission is generally ascribed to a region of high density, filled with a magnetic field and moving in a relativistic jet aligned to the line-of-sight. \\

While there is a general consensus that the low energy bump is due to synchrotron emission from electons and positrons in the emitting region, the origin of the high energy emission is interpreted differentely in leptonic or hadronic models. In the first scenario, the emission is entirely dominated by leptons, which produce the high energy bump through inverse Compton scattering off the synchrotron photons themselves (Synchrotron-Self-Compton, SSC) or off an external photon field. In the hadronic scenario the high energy bump is usually attributed entirely to proton synchrotron emission or photo-hadronic interactions.\\

We have developped a lepto-hadronic code that reproduces both ``extreme'' scenarios and also allows the study of mixed scenarios with characteristic signatures in the high energy band.
  
\section{Code development}

\subsection{ Leptonic processes}
 
The framework of our model is the stationary one-zone SSC code developped by \citet{Kata01}. A spherical region (characterised by its radius $R$) moves in the relativistic jet with Doppler factor $\delta$ : it is filled with a homogenous magnetic field $B$ and a primary electron population n$_e$($\gamma$=$E/(mc^2)$), which is parametrised by a broken power law (defined by the two slopes $\alpha_{1,2}$, the Lorentz factors $\gamma_{e,min}$\  , $\gamma_{e,break}$\ , $\gamma_{e,max}$ and the normalization factor $K_e$ at $\gamma=1$).\\

The leptonic part of the original code has been improved as follows :\\

\begin{itemize}
\item The synchrotron emission is evaluated computing the exact integration over pitch angles. The approximation used by \citet{Kata01} is valid only in the energy range of interest if the emitting particles are $e^\pm$.
\item The internal absorption due to $\gamma$-$\gamma$ pair production is evaluated using the parametrization for the cross-section given by \citet{Aha08}, instead of a $\delta$-function. The pair injection rate from $\gamma$-$\gamma$ interactions is computed following \citet{Aha83}.  A stationary distribution of secondary $e^\pm$ pairs ($1^{st}$ generation), taking into account synchrotron cooling, is evaluated following \citet{Susumu}. Synchrotron emission from the stationary pair distribution is evaluated as well.
\item The absorption induced by the extra-galactic background light (EBL) on TeV photons is evaluated using the EBL model by \citet{Franceschini}. 
\end{itemize}

\subsection{Hadronic processes}

The spherical emitting region is filled with a primary proton population $n_p(\gamma)$, parametrized by a power law with slope $\alpha_p$, normalization factor K$_p$ at $\gamma=1$ and minimal and maximal Lorentz factor $\gamma_{p,min}$ and $\gamma_{p,max}$. The proton synchrotron emission is corrected for internal $\gamma-\gamma$ absorption and the associated first generation pair spectrum is evaluated as described above.\\

In hadronic scenarios, an important role is played by photo-hadronic interactions. They have been computed using the Monte Carlo code \textit{SOPHIA} \citep{Anita}. \\
The target photon field is composed by the synchrotron radiation from primary electrons and from the protons themselves. The energy of the interacting protons is corrected for synchrotron losses following \citet{Anita}. The \textit{SOPHIA} code is then called for 10 sampled proton energies. The distributions of the generated particles ($\gamma, e^\pm, p, n, \nu_{e,\mu}, \hat{\nu}_{e,\mu}$) are summed and normalized to the number of protons in the blob suffering $p$-$\gamma$ interactions.\\
The spectra of the generated e$^\pm$ are corrected for radiative cooling to arrive at a steady-state solution. Synchrotron emission from secondary leptons and from the first generation pair spectrum  from $\gamma$-$\gamma$ interactions is then evaluated. \\    
$\mu^\pm$ can emit synchrotron radiation before decaying into $e^\pm$. They are retrieved from \textit{SOPHIA}, and, in a first approximation, their synchrotron emission is evaluated without considering the competition between synchrotron photon emission and decay into $e^\pm$.

\section{Application to PKS 2155-304 data}
\begin{figure}[h!]
 \centering
 \includegraphics[width=1.05\textwidth,clip]{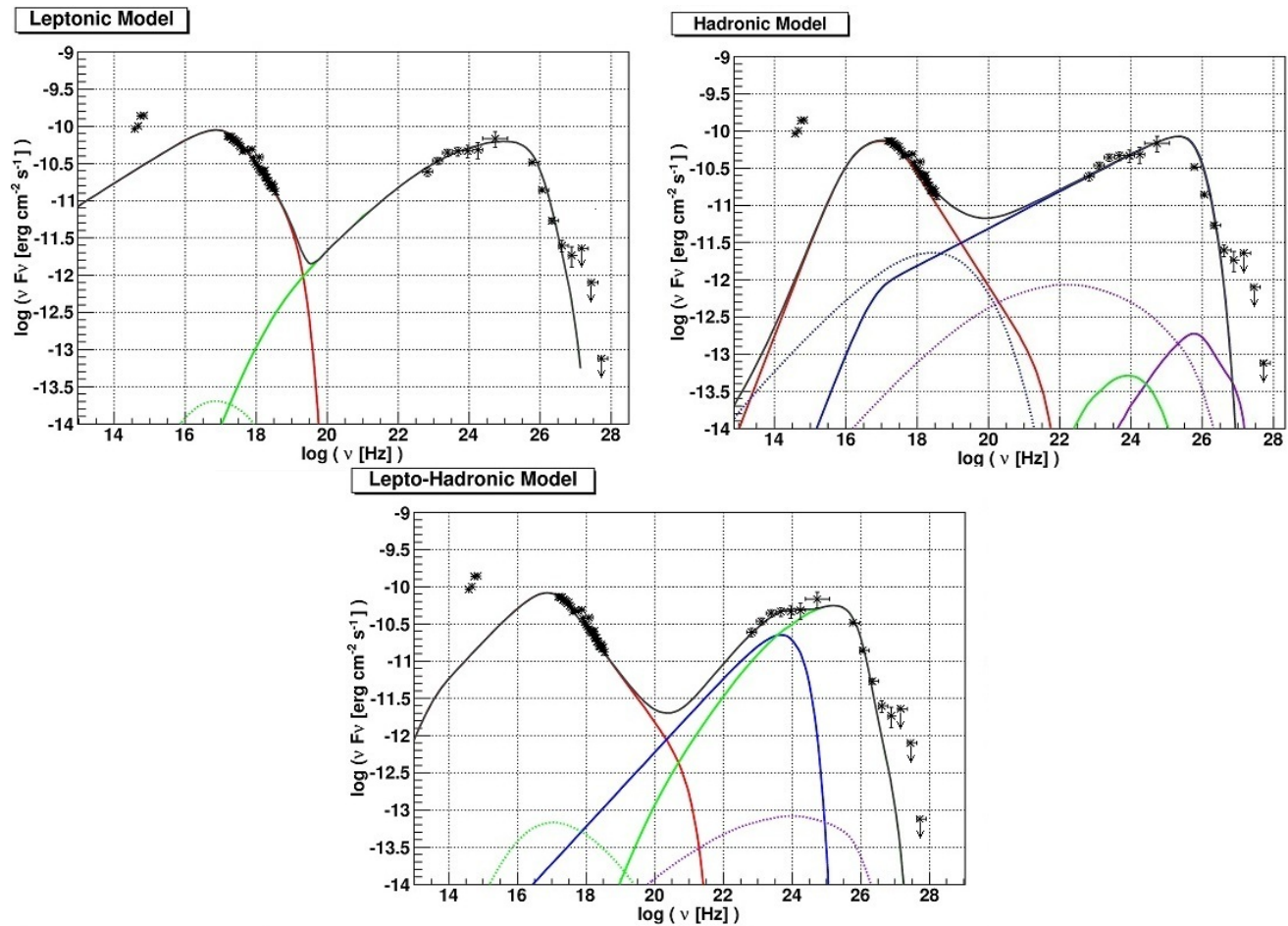}      
  \caption{Modelling of PKS2155-304 in three different scenarios. Colour code is as follow : primary electron synchrotron emission - red; inverse Compton emission - green; proton synchrotron emission - blue; muon synchrotron emission - violet. Dotted lines represent the synchrotron emission from the secondary pairs from $\gamma$-$\gamma$ pair production. \textit{Top left}: pure SSC model; \textit{top right}: proton synchrotron dominated model; \textit{bottom} : mixed lepto-hadronic scenario.}
  \label{fig1}
\end{figure}

As a first test of our code, we have reproduced the SED of a bright high frequency peaked blazar for different scenarios. Fig. \ref{fig1} shows an application to the data from the 2008 multiwavelength campaign on the blazar PKS 2155-304 \citep{PKS}. This data set represents at this time one of the most complete simultaneous SEDs of a TeV blazar: the shape of the high energy bump is fully characterised by the Fermi and H.E.S.S. data. PKS 2155-304 has been found to be in a low activity state during the campaign. In 
the interpretations presented here, the emission from the blob explains only the X-ray to $\gamma$-ray emission, assuming that from radio to visible light the emission from the extended jet dominates.\\

We report the modelling of the SED in three different scenarios : a pure SSC leptonic model, a proton synchrotron dominated hadronic model and a third, mixed, scenario in which both the synchrotron emission from protons and the inverse Compton component contribute to the high energy bump.\\

The modelling of the SED for the three different cases is shown in Fig. \ref{fig1}, and the model parameters are given in Table \ref{tab1}. The first scenario (top left in Fig. 1), is a standard SSC model, assuming a low magnetic field (75 mG), far from equipartition; the synchrotron emission from secondary particles from $\gamma$-$\gamma$ pair production is more than three order of magnitude below the primary synchrotron emission. The second scenario (top right in Fig. 1) is a proton synchrotron dominated hadronic model, in which the inverse Compton component is almost three order of magnitude lower than the proton synchrotron emission; $\mu$-cascade emission contributes (even though at $\sim1\%$ only) to the high energy bump; the magnetic field value is much higher (80 G) than in the SSC case, in order to have a significant contribution from protons; the presence of protons significantly increases the luminosity of the jet (equal to $10^{46}$  $\textrm{ergs s}^{-1}$) . The last case (bottom in Fig. 1) is an example of a mixed lepto-hadronic scenario, in which both inverse Compton and proton synchrotron emission contribute to the high energy bump; the magnetic field value is small (0.22 G), but the jet luminosity is high, given the high value of the emitting region size (ten times the value assumed for the proton synchrotron model).\\

%


\section{Conclusions}
The code presented here permits to reproduce the observed blazar SED in different scenarios, spanning a wide parameter space: a leptonic or a hadronic solution can be found assuming different physical conditions (magnetic field, density and distribution of primary particles) in the blob. In addition, mixed lepto-hadronic scenarios naturally arise in this framework.\\

 The increasing data quality at high energies, provided by {\it Fermi} in the GeV range, and by current and planned Cherenkov telescopes in the TeV range, will hopefully help to evaluate contributions from both leptons and hadrons. The next step towards a more realistic emission model would be a time-dependent lepto-hadronic code, which could take into account constraints from the observed variability from blazars. \\

\newpage
\begin{table}
	\centering
		\begin{tabular}{c c c c}
		\hline
		\hline
		& SSC & Proton synch & Mixed \\
		\hline
		$\theta$ & $1^\circ$ & $1^\circ$& $1^\circ$\\
		$\delta$ & $30$ & $30$ & $30$\\
		$\gamma_{e,min}$& $5\times10^2$ & $1\times10^3$ & $1\times10^3$ \\
		$\gamma_{e,break}$& $1.2\times10^5$ & $4\times10^3$ & $6\times10^4$\\
		$\gamma_{e,max}$& $1\times10^6$& $5\times10^5$&$5\times10^6$\\
		$\alpha_{e,1}$ &$2.4$&$2.2$ &$2.0$ \\
		$\alpha_{e,2}$ &$4.32$&$4.5$ &$4.25$ \\
		$K_e$ & $7\times10^4$ & $3.1\times10^3$ & $6.7\times10^3$\\
		$u_e$ & $5.4\times10^{-3}$ & $4.8\times10^{-4}$& $1.3\times10^{-2}$\\
		$\gamma_{p,min}$& -& $1\times10^5$ & $1\times10^5$\\
		$\gamma_{p,max}$& -& $5\times10^9$& $1\times10^{10}$\\
		$\alpha_{p}$ & -& $2.5$& $2.0$\\
		$\eta = K_p/K_e$ & -& $4.8\times10^4$& 8\\
		$u_p$ & -& $1.4\times10^3$&$9.2\times10^{2}$\\
		$R_{src}$ & $1.7\times10^{16}$& $5.0\times10^{14}$ & $4.8\times10^{15}$ \\
		$B$ & $0.075$& $80$ & $0.22$\\
		$u_B$ & $2.2\times10^{-4}$& $2.5\times10^2$& $2\times10^{-3}$\\
		$L_{jet}$& $5\times10^{43}$& $1\times10^{46}$ & $5\times10^{47}$\\
		\hline
		\hline			
		\end{tabular}
\caption{The table shows the different parameters used in the modelling of the PKS 2155-304 SED for the three different scenarios (SSC, synchrotron proton and mixed scenario) plotted in Figs. 1 to 3. Common values of redshift $z=0.116$, Doppler factor $\delta=30$ and viewing angle $\theta = 1^\circ$ have been used. The normalization parameter $K_{e,p}$ is in units of $\textrm{cm}^{-3}$, and represents the number density of the primary particle distribution at $\gamma = 1$; the size of the emitting ragion $R_{src}$ is in cm; the magnetic field in gauss. The energy densities $u_{e,p,B}$ are given in $\textrm{ergs cm}^{-3}$; the jet luminosity is in $\textrm{ergs s}^{-1}$. In the evaluation of the jet luminosity the cold proton content of the jet has been included (following \citet{Sikora}).}
\label{tab1}
\end{table}

\begin{acknowledgements}
The authors wish to thank Anita Reimer for providing the latest version of the SOPHIA code , and Hélène Sol for very useful discussions.
\end{acknowledgements}

\bibliographystyle{aa}  
\bibliography{cerruti} 

\end{document}